# TERS-ABNet: A Deep Learning Approach for Automated Single-Molecule Structure Reconstruction with Atomic Precision from TERS Mapping

Jie Cui[1,2], Yao Zhang[1,2,3]*, Yang Zhang[1,2,3], Yi Luo[1,2,3]*, Zhen-Chao Dong[1,2,3]*

[1]Hefei National Research Center for Physical Sciences at the Microscale and CAS Center for Excellence in Quantum Information and Quantum Physics, University of Science and Technology of China, Hefei, Anhui 230026, China

[2]Hefei National Laboratory, University of Science and Technology of China, Hefei, 230088, China

[3]School of Physical Sciences and Department of Chemical Physics, University of Science and Technology of China, Hefei, Anhui 230026, China

*Corresponding authors. Emails: zhy2008@ustc.edu.cn (Y.Z.); yiluo@ustc.edu.cn (Y.L.); zcdong@ustc.edu.cn (Z.C.D.)

## Abstract

Determining the chemical structure for a single molecule on surface from spectroscopic data represents a challenging high-dimensional inverse problem. Tip-enhanced Raman spectroscopy (TERS) enables chemically specific imaging of single molecules with sub-nanometer spatial resolution, yet reconstructing complete molecular structures from TERS maps remains difficult owing to the ambiguous vibrational signatures and reliance on expert interpretation. Here, we introduce TERS-ABNet, a deep-learning framework that formulates single-molecule structure determination from spectroscopic images as an image-to-graph inference task. Using a "two-track" architecture, the model jointly predicts probabilistic atom and bond maps, enabling direct construction of explicit atom-bond graphs without relying on predefined chemical rules. Trained on simulated datasets, TERS-ABNet achieves ~94% atom-type classification accuracy (with a mean coordinate error of ~0.23 Å), enabling to reliably recovering molecular connectivity and fully reconstruct single-molecule structure from its TERS maps. The framework generalizes across varying spatial resolutions and structural complexity through transfer learning, and successfully reconstructs the atomic structure of a single porphyrin molecule from experimental TERS data. This work establishes a general deep-learning strategy for inferring explicit atom-bond graph representations from high-dimensional spectroscopic imaging data, providing a new pathway towards automated molecular structure determination in nanoscale characterization.

## Introduction

Inferring the atomic structure of single molecules on surfaces from experimental measurements represents a long-standing challenge in nanoscale science. Reliable access to single-molecule structural information is essential for understanding surface-mediated chemistry[1,2], molecular self-assembly[3], and the function of molecular nanodevices[4]. Scanning probe microscopies, including scanning tunneling microscopy (STM)[5–7] and atomic force microscopy (AFM)[7–9], provides sub-nanometer real-space imaging that can reveal molecular skeletons and even individual atom positions. However, the limited chemical contrast of these techniques, together with tip-induced perturbations and substrate coupling, often prevents unambiguous identification of element types and bonding connectivity. Consequently, determining the complete structure of a single molecule at single-atom level from experimental imaging data remains a great challenge.

Tip-enhanced Raman spectroscopy (TERS) has emerged as a powerful spectroscopic imaging technique that combines sub-nanometer spatial resolution with chemically specific vibrational fingerprints[10–15], enabling visualization of single molecules and, in favorable cases, even chemical bonds[12,13]. However, translating TERS measurements into complete molecular structures remains a high non-trivial inverse task[16]. The single-molecule Raman response in strongly confined plasmonic fields exhibits complex polarization- and orientation-dependent behavior[17,18], leading to complex spectral patterns that obscure direct correspondence between local imaging features and atomistic structure. Consequently, structural interpretation from TERS data still relies heavily on expert-driven vibrational assignments and chemical intuition, limiting scalability and robustness for structurally complex motifs (such as conjugated and heterocyclic systems)[13].

More broadly, reconstructing explicit atomistic graph representations from high-dimensional spectroscopic imaging constitutes a difficult inverse problem that has not yet been solved in an automated and generalizable manner. Recent advances in deep learning have enabled data-driven modeling of structure–property[19,20] and structure–spectroscopic relationships[21–23], yet these approaches typically produce ensembles of candidate structures rather than unique atomistic solutions[23,24], reflecting the degeneracy of conventional spectroscopic descriptors[25]. In parallel, machine learning applied to STM and AFM has achieved impressive performance in predicting molecular skeletons and atomic positions from real-space images[26–28], but often lacks the chemical specificity required for unambiguous structural determination. Therefore, bridging chemically rich spectroscopic information with spatially resolved atom-bond graphs represents an open machine-intelligence challenge in nanoscale characterization.

Here, we present TERS-ABNet, a deep-learning framework that formulates single-molecule structure determination from TERS images as a structured atom-bond inference problem. This framework integrates an atom-prediction network (ANet) and a bond-prediction network (BNet) to jointly predict atomic positions, identify atomic species and infer chemical connectivity directly from spectroscopic maps. For planar molecules, TERS-ABNet provides sufficient and precise constraints to reconstruct complete molecular structures, accurately resolving atomic positions and interatomic bonds, including complex conjugated and heterocyclic ring motifs that are difficult to resolve using conventional intuition-driven analysis[13,29]. This framework generalizes across varying spatial resolutions from Ånström to sub-nm level, and atomic-level identification can be always computationally recovered from corresponding TERS maps, relaxing stringent experimental requirements on spatial resolution. Through transfer learning, TERS-ABNet further generalizes to nonplanar and three-dimensional molecules, demonstrating robustness to orientation-dependent selection rules and incomplete spectroscopic sampling. We validate the practical applicability of this approach by reconstructing of atomic positions and bond connectivity of a single porphyrin molecule from experimental TERS data. This work establishes a general machine-learning paradigm for inferring molecular atom-bond graph representations from spatially and chemically resolved spectroscopic imaging, paving the way toward automated structure determination across scanning-probe-based spectroscopic platforms and highlighting the potential of data-driven solutions to long-standing inverse problems in nanoscale characterization.

## Results and discussions

**Theoretical description of single-molecule TERS as a structured inverse problem**

Figure 1 illustrates the physical basis of single-molecule TERS using 2,6-dihydroxypyridine molecule as a representative example, serving to demonstrate how atomic positions and interatomic connections can, in principle, be inferred from spatially resolved spectroscopic images. The discussion below is general and applies to arbitrary molecular systems. Owing to the extreme spatial confinement of plasmonic fields in a nanocavity[30] and the position-dependent polarization response of a single molecule[17,18,31], the measured TERS signals depends sensitively on the tip position. As a result, scanning the tip across the molecule enables spatial mapping of individual vibrational modes (Fig. 1a). Within established theoretical descriptions of single-molecule TERS in plasmonic nanocavities[32,33], the Raman intensity ($I_k$) of the $k$-th vibrational mode ($Q_k$) measured with the tip located at position $\mathbf{R}_0$ can be expressed as

$$I_k(\mathbf{R}_0) \propto \left| \int \vec{\mathbf{G}}_k(\mathbf{r}_\infty, \mathbf{r} - \mathbf{R}_0; \omega_k) \mathbf{p}_k^{\text{loc}}(\mathbf{r}, \mathbf{R}_0) d\mathbf{r} \right|^2, \tag{1}$$

where $\vec{\mathbf{G}}_k(\mathbf{r}_\infty, \mathbf{r} - \mathbf{R}_0; \omega_k)$ is the far-field dyadic Green's function associated with the molecular Raman dipole moment $\mathbf{p}_k$ at radiation energy $\hbar\omega_k$. In the representation of atomic-orbital basis, the molecular Raman dipole moment associated with vibrational mode $Q_k$ can be written as[13]

$$\mathbf{p}_k^{\text{loc}}(\mathbf{r},\mathbf{R}_0) = \sum_{\alpha}\sum_{\beta} C_\alpha^* C_\beta\, g(\mathbf{r}-\mathbf{R}_0) \times \left[\varphi_\alpha^*(Q_k,\mathbf{r})\,\mathbf{r}\,\varphi_\beta(Q_k,\mathbf{r})\right], \qquad (2)$$

where $g(\mathbf{r} - \mathbf{R}_0)$ describes the spatial distribution of the confined plasmonic field centered at the tip position $\mathbf{R}_0$, $\varphi_{\alpha(\beta)}$ denotes atomic orbitals corresponding to atom $\alpha(\beta)$, and $C_{\alpha(\beta)}$ represents the coefficient for corresponding electronic states. Because both atomic orbitals and nanocavity plasmonic fields are spatially extended, the resulting Raman intensity reflects coherent and interferential contributions from multiple neighboring atoms and chemical bonds (Fig. 1b, see Supplementary Section 1 for details). Consequently, a single TERS image of an individual vibrational mode does not correspond uniquely to a specific atom or bond, rendering direct interpretation of TERS maps in terms of molecular structure highly non-trivial. Nevertheless, when a sufficiently rich set of TERS images spanning multiple vibrational modes is acquired, each mode provides a distinct projection of the underlying atomic and bonding configuration. In principle, the complete molecular structure is therefore implicitly encoded across a high-dimensional set of spectroscopic images, giving rise to a structured inverse problem: inferring discrete atomic positions and bond connectivity from continuous, spatially overlapping spectroscopic images.

**Model framework and data representation**

To address this inverse problem, we developed TERS-ABNet, a deep-learning framework that reconstructs molecular structure by jointly inferring atomic positions, atomic species and chemical bonds directly from TERS mapping images. As schematically illustrated in Figs. 1c–1j, TERS-ABNet, adopts a "two-track" architecture consisting of an atom-prediction network (ANet) with a bond-prediction network (BNet). Both networks share the same underlying encoder-decoder architecture based on two-dimensional attention U-Nets[34,35] (Fig. 1k, see Methods and Supplementary Section 2 for details), enabling consistent feature extraction from identical spectroscopic inputs while producing complementary structured outputs. Each network comprises 12 convolutional layers and is designed to simultaneously reduce the spectral dimensionality of the input TERS data and disentangle the spatial contributions associated with individual atoms or chemical bonds. Three attention gates are incorporated through skip connections to selectively emphasize spatial−spectral features relevant for atom or bond identification. The ANet and BNet are trained to predict multi-channel probability maps corresponding to predefined atomic species and bond types, respectively (Figs. 1e and 1h). A subsequent spot-detection algorithm is applied to these maps to extract the coordinates of atoms and bonds (Figs. 1f and 1i). Based on the predicted coordinates and types, atomic connectivity is then determined algorithmically, yielding a complete molecular structure in an automated manner (Fig. 1j).

The training of TERS-ABNet requires paired datasets consisting of simulated TERS mapping images and corresponding structured ground-truth representations of atomic or bonding information for individual molecules. In total, 23,284 samples were generated through data augmentation of 5823 molecular structures (see Methods and Supplementary Section 3 for details). For each molecule, TERS spectra were simulated on a 128 × 128 spatial grids spanning an area of 2.5 × 2.5 nm$^2$, ensuring complete coverage of the molecular geometry. To construct structured input for the network, the simulated spectra were integrated over 20 cm$^{-1}$ wavenumber windows, yielding 160-channel TERS mapping image of 128 × 128 pixels. The corresponding ground-truth outputs were encoded as multi-channel two-dimensional images in which atoms or bonds were represented as Gaussian-shaped spots centered at their true positions. Atomic labels were encoded using four channels corresponding to C, N, O and H atoms, whereas chemical bonds were represented using nine channels corresponding to predefined functional bond types (see Supplementary Section 3 for details). This explicit spatial–spectral representation enables TERS-ABNet to learn a direct mapping from high-dimensional spectroscopic images to structured atom–bond images, forming the inference foundation for automated single-molecule structure reconstruction.

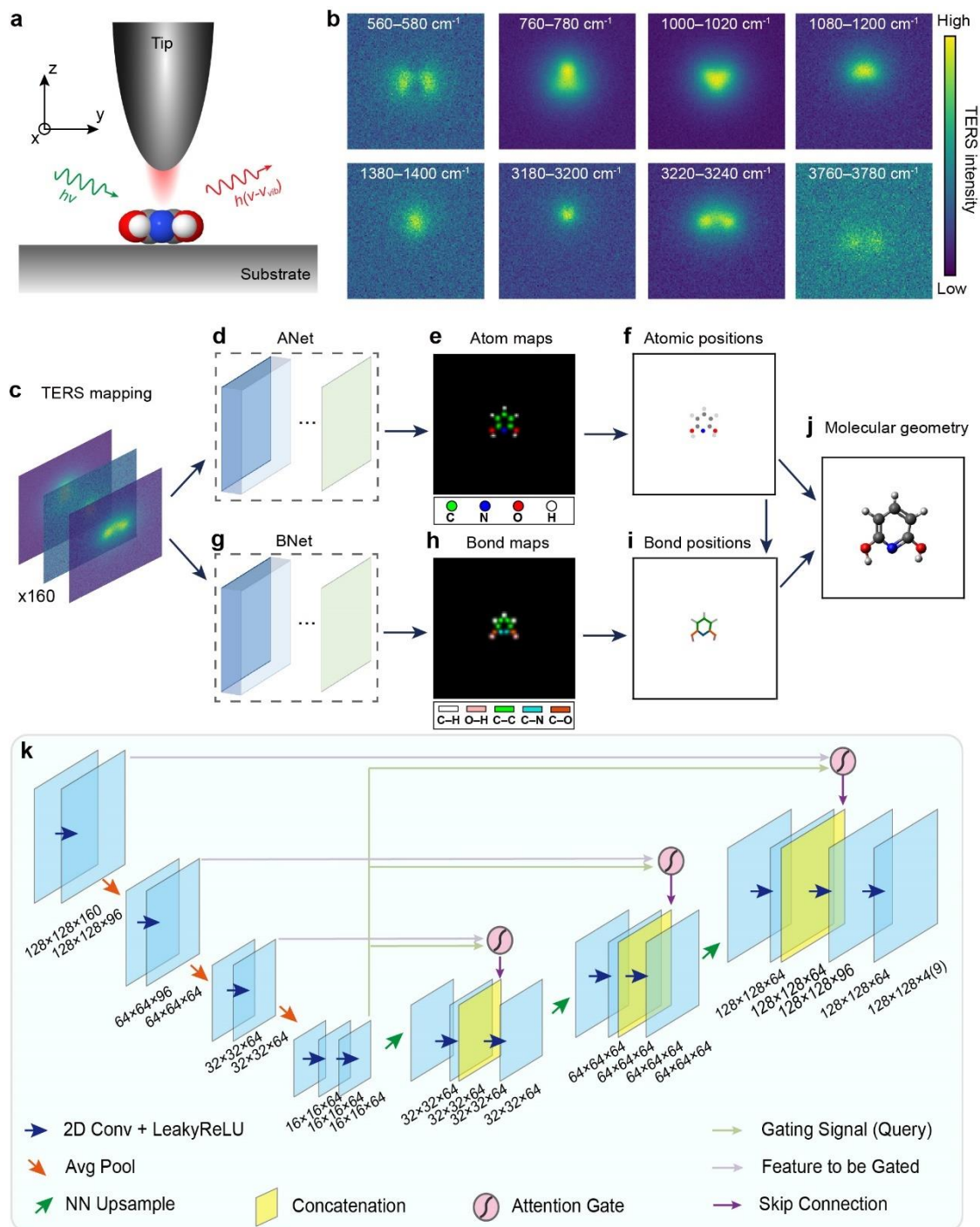

**Fig. 1. Workflow of TERS-ABNet for atom and bond identification and molecular structure reconstruction from TERS mapping. a**, Schematic of TERS setup, in which a single molecule positioned under a metal tip apex generates Raman signals under laser illumination. **b**, Representative TERS mapping images simulated within selected wavenumber windows, showing distinct spatial distributions associated with different vibrational modes. **c**, **d**, **e**, Workflow for atom prediction: multi-channel TERS maps (**c**) are processed by the atom-prediction network (ANet, **d**) to generate element-

resolved atom probability maps (**e**). **c**, **g**, **h**, Workflow for bond prediction: the same TERS maps (**c**) are input into the bond-prediction network (BNet, **g**) to produce bond-type-resolved probability maps (**h**). **e**, **f**, **h**, **i,** Atomic (**f**) and bond (**i**) coordinates are extracted from the predicted probability maps (**e**, **h**) using a spot-detecting algorithm. **j**, The molecular geometry reconstructed by combining the detected atomic positions (**f**) with bond connectivity inferred from the predicted bonds (**i**). **k**, Architecture of ANet (and BNet), based on a two-dimensional attention U-Net. The dimensions at each layer are indicated as Height × Width × Channel (H×W×C), where H×W denotes the spatial extension and C represents the spectral-channel dimension of the TERS input. Size of image: 2.5×2.5 nm$^2$. Complete sets of TERS maps, corresponding ground-truth atom and bond maps, and molecular geometries are provided in Supplementary Fig. 3.

**Accuracy of atom and bond prediction**

We first quantitatively evaluate the performance of TERS-ABNet in predicting atomic positions, atomic species and chemical bonds using an independent test set. For each test molecule, the trained networks generate atom and bond probability maps, from which the coordinates of atoms and bonds are extracted using a spot-detection algorithm[36]. Predicted atoms and bonds are then matched to their ground-truth counterparts via a nearest-neighbor matching algorithm[37], enabling systematic evaluation of both position accuracy and type classification. Figures 2a and 2b compares the predicted and true coordinates of atoms and chemical bonds, respectively. The mean absolute error (MAE) of the predicted atomic coordinates is ~0.233 Å, with 98.2% of atoms localized within this error tolerance, while the MAE for bond positions is ~0.199 Å with an accuracy of 98.2%. These localization errors correspond to only approximately 10−15% of a typical covalent bond length (~1.5 Å) [38], demonstrating that TERS-ABNet achieves highly precise spatial inference of both atoms and bonds from spectroscopic images alone.

Figures 2c and 2d show the confusion matrices for atom-type and bond-type classification. The mean classification accuracy reaches 94.2% for atomic species and 92.0% for chemical bonds, respectively. Hydrogen atoms exhibit the highest prediction accuracy, which can be attributed to their terminal positions within molecular structure and the relatively localized vibrational signatures associated with X−H (where X can be C, N, or O elements) stretching modes. In contrast, nitrogen and oxygen atoms are predicted with slightly lower accuracy (88.4% and 89.3%, respectively; Fig. 2c), likely because these heteroatoms are often embedded within molecular backbones and lack uniquely identifiable vibrational fingerprints in TERS images. Notably, the bond-prediction network (BNet) reliably distinguishes hydrogen-containing terminal bonds, such as C−H, N−H, $NH_2$ and O−H bonds (Fig. 2d). Bonds located within the molecular backbone (including C–N, C–O, N−N, and N–O bonds) exhibit comparatively lower classification accuracy, reflecting the increased spectral overlap and interferential contributions from neighboring atoms in these regions. These results highlight the intrinsic difficulty of bond-level inference in densely connected molecular motifs and underscore the importance of jointly leveraging spatial and spectral information.

We further analyze structural completeness by quantifying missing and spurious predictions. The inherent selection rules of TERS and spectral broadening effects can suppress certain vibrational modes, leading to missing atoms or bonds in the predicted structures, whereas experimental noise may introduce spurious atoms or bonds in the predictions. As shown in Figs. 2e and 2f, instances with extra atoms or bonds occur less frequently than those with missing entities. In the test set (n = 2332), 931 molecules predicted by ANet exhibit neither missing nor extra atoms, while 685 molecules predicted by BNet were reconstructed without bond insertion or deletion errors. The overall reconstruction accuracy achieved here substantially exceeds that of conventional structure elucidation approaches based solely on one-dimensional infrared or Raman spectra[23], for which top-ranked structure prediction accuracies are typically limited to only a few percent owing to severe information loss and spectral overlap inherent in ensemble-averaged and orientational-averaged spectroscopic descriptors[25]. By contrast, TERS-ABNet leverages spatially resolved spectroscopic images, effectively lifting this degeneracy and enabling atom- and bond-resolved structure inference with improvements of several orders of magnitude in accuracy. This comparison underscores the critical role of incorporating spatial information into spectroscopic learning frameworks for reliable molecular structure determination.

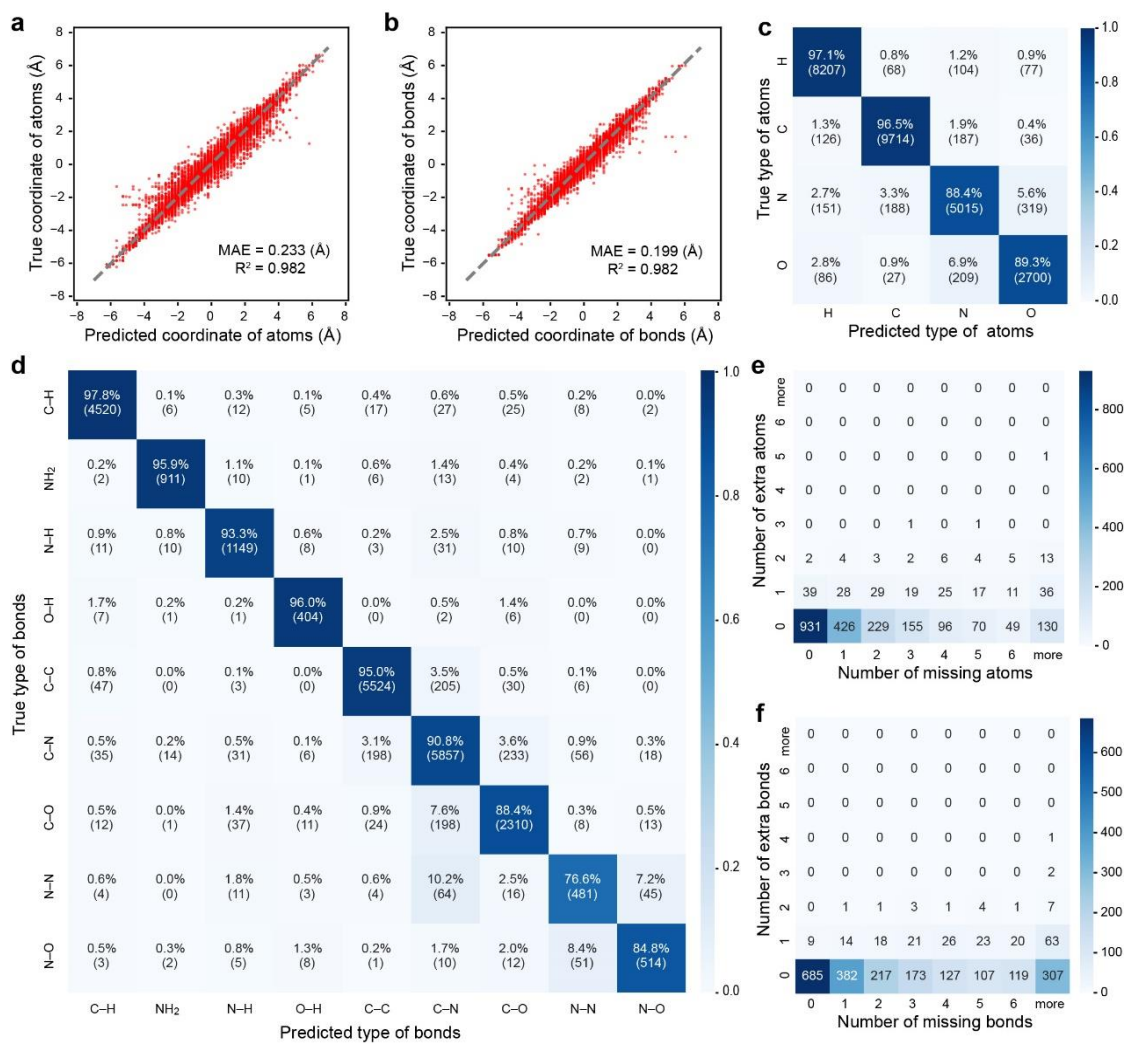

**Fig. 2. Quantitative evaluation of atom and bond prediction accuracy. a**, Regression analysis comparing ANet-predicted atomic coordinates with ground-truth positions. **b**, Regression analysis comparing BNet predicted bond coordinates with ground-truth positions. **c**, Confusion matrix for atomic type classification by ANet. **d**, Confusion matrix for bond-type classification by BNet. Values in **c** and **d** are normalized by the total number of samples in each ground-truth class. **e**, Distribution of molecules with varying numbers of missing and extra atoms predicted by ANet. **f**, Distribution of molecules with varying numbers of missing and extra bonds predicted by BNet.

**Geometry construction of molecules from simulated TERS mappings**

The high accuracy achieved in atom and bond prediction enables TERS-ABNet to reconstruct complete molecular geometries directly from simulated TERS mappings. To demonstrate this capability, we selected four representative planar molecules from the test set that encompass a range of structural motifs, including short-chain backbones, conjugated aromatic rings, heterocycles and fused-ring systems. These examples collectively probe the ability of the model to resolve increasing levels of chemical and topological complexity. As a general observation, the far-field radiation component of individual TERS maps does not exhibit features that can be readily interpreted at atomic resolution. Nevertheless, when integrated across multiple vibrational modes and processed by TERS-ABNet, the underlying atomic-scale structure becomes explicitly recoverable.

We first consider the short-chain molecule N'-methanimidoyl-2-oxoethanimidamide (Figs. 3a–3d). For this relatively simple system, the predicted atom and bond probability maps exhibit excellent agreement with the reference structure (Supplementary Fig. 4), yielding well-separated and sharply localized spots corresponding to all atoms and chemical bonds. The subsequent spot-detection and graph-construction procedures correctly identify all atomic species and recover the complete bonding network without ambiguity, resulting in a molecular geometry that is fully consistent with the ground truth.

We next examine the 3-ethynyl-1H-pyridin-2-one molecule, which contains a six-membered heteroaromatic ring (Figs. 3e–3h). The vibrational signatures associated with aromatic rings are notoriously difficult to disentangle using conventional spectroscopic interpretation owing to strong mode delocalization and similarities in Raman fingerprints[7,13]. Despite these challenges, TERS-ABNet accurately localizes all atomic positions and correctly classifies the bonds forming the pyridine ring, demonstrating its ability to infer conjugated ring geometries directly from spatially resolved spectroscopic data.

A more challenging example is provided by the fused heterocyclic system 1H-cyclopenta[d]pyrimidine, which incorporates both five-membered and six-membered heterocycles within a single molecule (Figs. 3i–3l). In this case, TERS-ABNet successfully predicts all atomic species and recovers the full bonding topology across the fused rings, underscoring its robustness in handling polycyclic and chemically intricate planar systems.

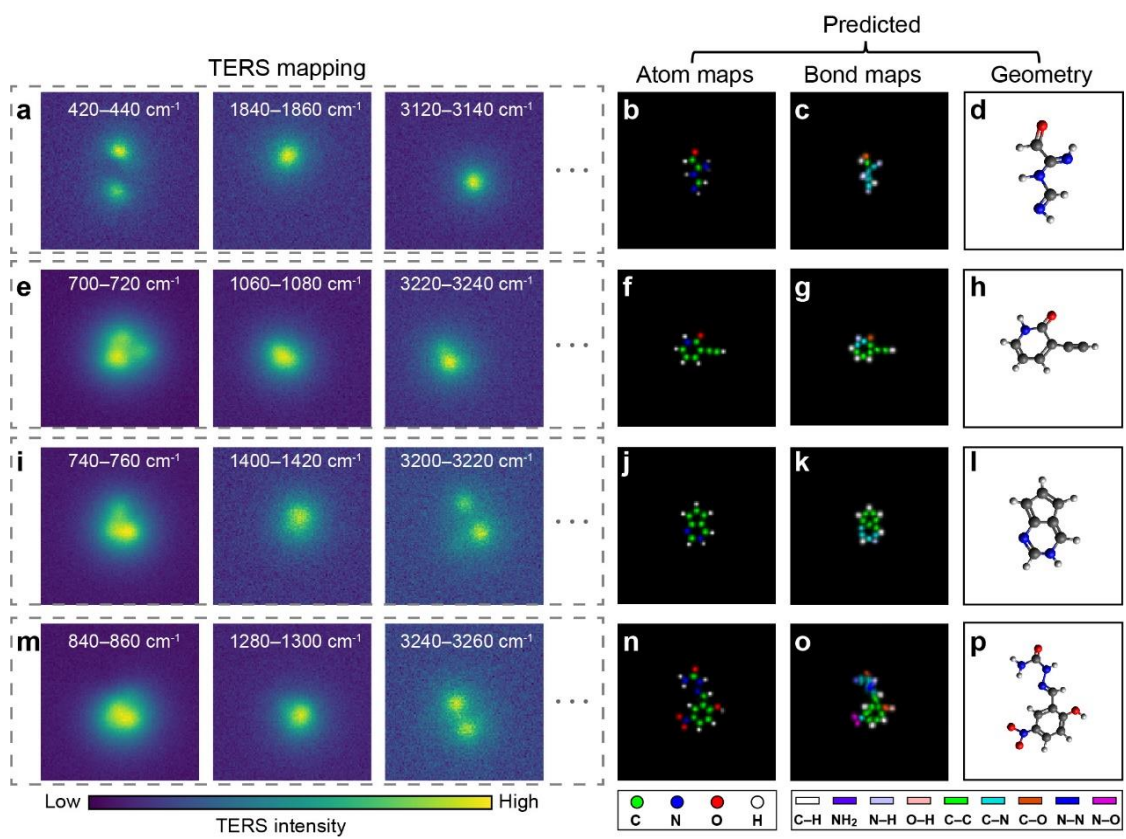

**Fig. 3. Reconstruction of representative molecules from simulated TERS mappings. a−d,** Short-chain molecule N'-methanimidoyl-2-oxoethanimidamide. **e−h,** Six-membered heterocycle 3-ethynyl-1H-pyridin-2-one. **i−l,** Fused heterocyclic system 1H-cyclopenta[d]pyrimidine, containing both five- and six-membered rings. **m−p,** Larger branched molecule [(E)-(2-hydroxy-5-nitrophenyl)methylideneamino]urea. For each example, columns 1 to 3 show simulated TERS mappings in three representative wavenumber regions. Columns 4 to 5 show merged images from element-resolved atomic probability maps predicted by ANet (4 channels) and bond-type-resolved probability maps predicted by BNet (9-channels), respectively. Column 6 presents the reconstructed molecular geometries obtained by integrating the predicted atomic positions with bond connectivity inferred from predicted chemical bonds. Size of image: 2.5 × 2.5 nm$^2$. Ground truth atom maps, bonds maps and molecular geometries are provided in Supplementary Fig. 4.

Finally, we consider [(E)-(2-hydroxy-5-nitrophenyl)methylideneamino]urea, a molecule comprising a six-membered aromatic ring and multiple peripheral functional groups (Figs. 3m–3p). The atom-prediction network ANet localizes all atoms with correct elemental classification, while the bond-prediction network BNet accurately resolves the majority of chemical bonds, including that within the aromatic core and substituents. Although the dense arrangement of bonds leads to partial overlap in the predicted bond probability maps, the individual bonds remain distinguishable, allowing successful reconstruction of the complete molecular geometry. The resulting structure is in full agreement with the reference, demonstrating that TERS-ABNet remains effective even in complex planar molecules.

**Generalization across varying TERS spatial resolutions**

To assess the robustness of TERS-ABNet under experimentally relevant conditions, we perform zero-shot tests across a range of spatial resolutions by varying the tip–molecule distance (Fig. 4 and Supplementary Section 5). As mentioned above, this model is trained using TERS mapping images simulated with a characteristic local-field spatial resolution of 0.87 nm. When the resolution is increased to 0.47 nm (corresponding to shorter tip–molecule distance and stronger near-field confinement), TERS-ABNet continues to precisely localize most atomic positions and bonding features, with only minor degradation in the recovery of subtle structural details.

More importantly, even when the spatial resolution is substantially reduced to 1.12 nm (representing more weakly confined and experimentally accessible measurement conditions with larger tip–molecule distance), the model remains capable of reliably capturing the overall molecular skeleton and principal connectivity patterns. Although the predicted atom- and bond-probability distributions become spatially broadened and may include limited spurious assignments, the essential structural topology is still correctly inferred. This behavior indicates that the network can effectively integrate spatially delocalized spectro-microscopic signals with learned structural priors, enabling the extraction of atomistic information from incomplete or low-resolution measurements.

In conventional TERS-based structure determination, achieving Ångström-scale field confinement is often regarded as a prerequisite for resolving atomic features. Our results indicate that this requirement can be partially alleviated by leveraging statistical correlations between spatially distributed spectroscopic signals and underlying molecular geometry. In this framework, the reconstruction accuracy is no longer determined solely by the intrinsic optical resolution, but also by the information content embedded in multimode vibrational responses and the model's ability to integrate such information into a coherent structural representation. This perspective suggests a shift in the role of machine learning in nanoscale spectroscopy: from merely enhancing image quality to fundamentally augmenting the effective information bandwidth of experimental measurements. By enabling reliable recovery of molecular frameworks under moderate near-field confinement, the approach opens new opportunities for high-throughput and experimentally accessible single-molecule characterization. More generally, it highlights the potential of data-driven inference strategies to mitigate physical measurement constraints and to extend the practical limits of structure determination in complex nanoscale systems.

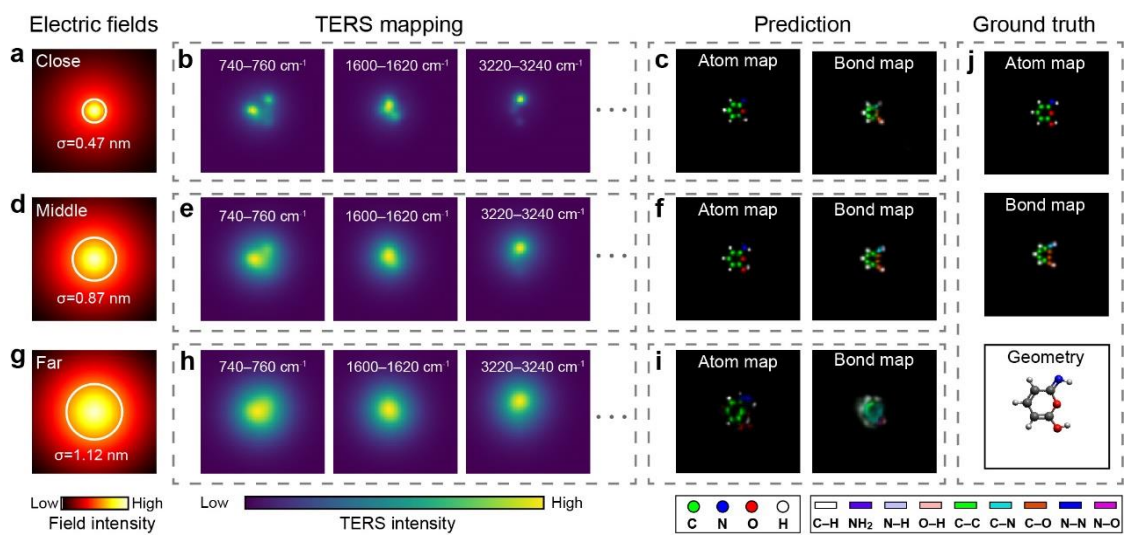

**Fig. 4. Generalization performance of the model across different TERS resolutions. a–c**, The case of "close" tip–molecule distance with $d = 0.1$ nm. **d–f**, The case of "middle" tip–molecule distance with $d = 0.2$ nm. **g–i**, The case of "far" tip–molecule distance with $d = 0.3$ nm. **j**, Ground truth of atom map, bond map, and molecular geometry. **a**, **d**, **g**, Spatial distributions of the local electric fields at the molecular plane for the "close" (**a**), "middle" (**d**) and "far" (**g**) tip−molecule distances. The white circles represent the location of the full width at half maximum (FWHM) of the local field to estimate the values of different σ. **b**, **e**, **h**, Simulated TERS mappings at representative wavenumber regions for the "close" (**b**), "middle" (**e**) and "far" (**h**) tip−molecule distances. **c**, **f**, **i**, Predicted atom map and bond map from the simulated TERS mapping at the "close" (**c**), "middle" (**f**) and "far" (**i**) tip−molecule distances. Image size: 2.5 × 2.5 nm$^2$.

**Generalization to complex molecules via transfer learning**

The preceding demonstrations focus exclusively on planar molecular systems. Whether the TERS-ABNet framework can be extended to nonplanar molecules remains a critical open question. It is important to note that TERS is intrinsically best suited for planar molecules or 2D systems on surface. For molecules containing out-of-plane functional groups, TERS measurements often suffer from incomplete vibrational mode sampling and reduced signal-to-noise ratios, owing to orientation-dependent selection rules and relatively large tip−molecule distance[39,40]. Nonetheless, to further assess the transferability and scalability of TERS-ABNet beyond strictly planar geometries, we constructed a dedicated dataset for transfer learning that includes molecules featuring methyl and ethyl substituents, which introduce localized three-dimensional structural motifs while preserving a largely planar molecular backbone. This setting provides a controlled testbed for probing nonplanar effects without departing entirely from the experimental regime in which TERS is most effective. For this transfer-learning task, the output layer of BNet was expanded from 9 to 11 channels, with the two additional channels corresponding to the methyl and ethyl groups, respectively. The architecture of ANet was left unchanged, as no new elemental species were introduced. The weights obtained from previous training on planar molecules were used to initialize the model, efficient reuse of learned spectra-spatial features while requiring only minimal retraining on the extended dataset.

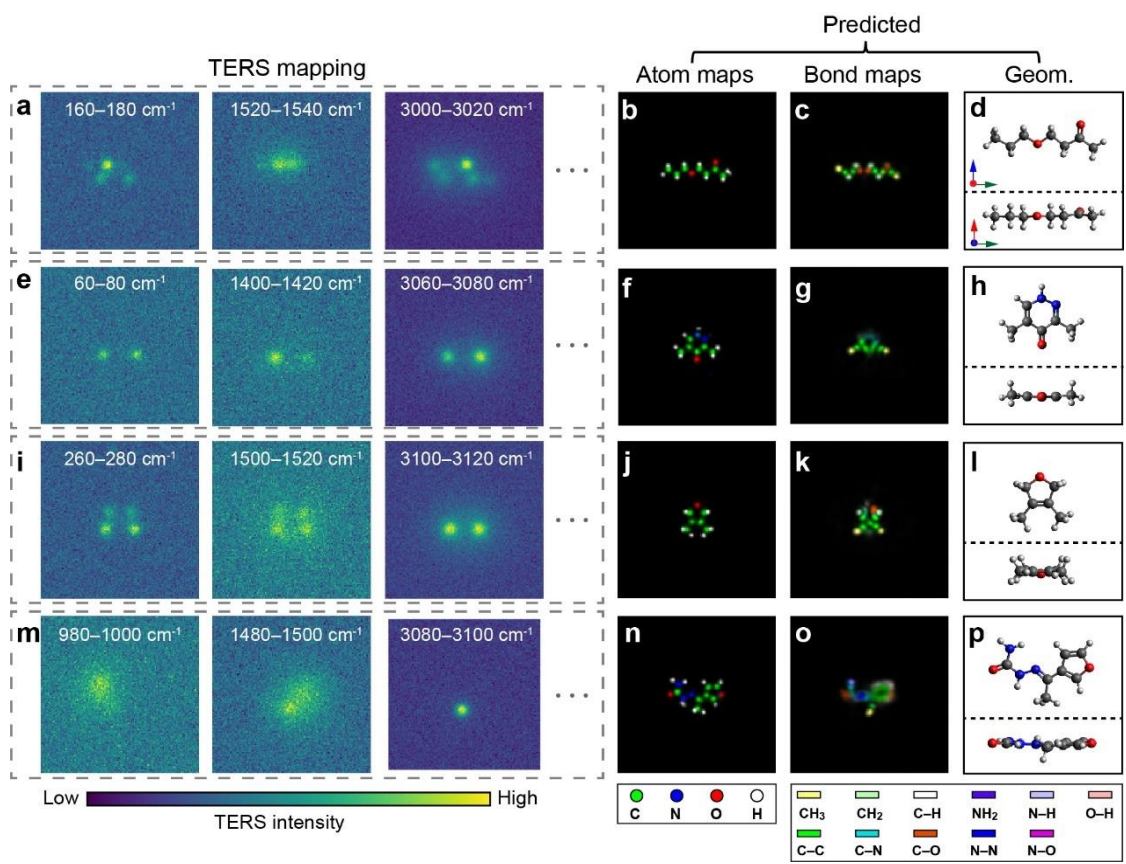

**Fig. 5. Reconstruction of representative nonplanar molecules from simulated TERS mappings. a−d**, A long-chain molecule 4-propoxybutan-2-one. **e−h**, Six-membered heterocycle 3,5-dimethyl-1H-pyridazin-4-one. **i−l**, Five-membered heterocycle 2,3,4,5-Tetramethylfuran. **m−p**, A structurally more complex molecule, [(E)-1-(furan-3-yl)ethylideneamino]urea. For each example, columns 1 to 3 present simulated TERS mappings in three representative wavenumber regions. Columns 4 to 5 show merged images from element-resolved atomic probability maps predicted by ANet (4 channels) and bond-type-resolved probability maps predicted by BNet (11 channels), respectively. Column 6 shows the reconstructed molecular geometries generated from the predicted atomic positions with bond connectivity from predicted bonds. To highlight their stereostructures for each molecule, both top views (upper panels) and front views (lower panels) are shown in column 6. Size of image: 2.5 × 2.5 nm$^2$. Ground truth atom maps, bonds maps and molecular geometries are provided in Supplementary Fig. 6.

To evaluate the performance of the transfer-learned model, we selected four representative molecules with diverse nonplanar substituents from the test set. As shown in Fig. 5 (columns 1 and 2), the presence of three-dimensional functional groups leads to a pronounced attenuation of Raman signals in the 0–2000 cm$^{-1}$ range, which is dominated by skeletal vibrational modes. Despite this substantial reduction in spectral contrast, both atomic and bonds predictions remain highly accurate. This result indicates that TERS-ABNet can robustly extract structurally relevant features from noisy and incomplete spectroscopic inputs, even in the presence of out-of-plane distortions. We note that for certain stereofunctional bonds, atoms located closer to the substrate may be partially obscured by upper atoms positioned nearer to the tip apex, resulting in reduced detection sensitivity at these sites. Nevertheless, BNet predicts methyl and ethyl groups with high fidelity, providing strong chemical constraints that enable the recovery of occluded hydrogen atoms during subsequent geometry construction. Moreover, even in cases where the predicted bond maps along the molecular backbone appear blurred (Fig. 5, column 5), interatomic distance criteria can be combined with the predicted atomic positions to infer bond connectivity. This complementary use of spatial and chemical information allows successful reconstruction of the overall molecular structure (Fig. 5, column 6).

We further explored the application of the transfer-learned model to more strongly three-dimensional molecules (Supplementary Section 7). In these cases, the reconstruction accuracy degrades markedly. This limitation does not arise from the learning framework itself, but rather from the intrinsic incompleteness of TERS mappings for fully three-dimensional molecules. Orientation-dependent selection rules and localized field confinement prevent a complete sampling of vibrational modes, resulting in insufficient information content to uniquely determine the full molecular geometry. These findings highlight a fundamental physical constraint on structure reconstruction from TERS data and delineate the current boundary of applicability for data-driven molecular geometry inference based on near-field vibrational spectroscopy.

**Geometry construction from experimental TERS mapping**

While the performance of TERS-ABNet has been systematically validated on simulation datasets, its practical relevance ultimately hinges on its applicability to experimentally acquired TERS measurements. To this end, we evaluated the model using experimental TERS mapping data of magnesium porphyrin (MgP) molecule, a prototypical conjugated macrocycle for which relatively complete experimental TERS mapping datasets are available[13]. As a reference, we first applied the trained model to the simulated TERS mapping images of MgP molecule. The simulated maps (Fig. 6a) were processed by ANet and BNet to generate atom-resolved (Fig. 6b) and bond-resolved (Fig. 6c) probability distributions. The predictions correctly localized all 12 hydrogen atoms and all peripheral carbon atoms, as well as the carbon atoms on the inner skeleton of the pyrrole rings. In contrast, the four nitrogen atoms coordinated to the central magnesium atom were misclassified as oxygen or carbon. This discrepancy likely arises from the strong spectral contribution associated with the central metal site, which perturbs the local vibrational response and introduces features not represented in the training data, as magnesium was not explicitly included as an atomic category in the model.

The predicted bond probability maps provide particularly informative insights. As shown in Fig. 6c, most of conjugated C−H bonds on the pyrrole rings and at the bridge sites are accurately identified as discrete and well-localized spots, yielding detailed positional information for these terminal C−H bonds. In contrast, C−C bonds are predicted as extended line-like features, while C−N bonds appear as a diffuse ring distributed inside the C−C framework. Notably, during training the bond labels are represented as localized Gaussian spots. The emergence of line- or ring-shaped predictions therefore reflects an intrinsic ambiguity in bond localization within large conjugated systems, where electronic states are delocalized over the entire macrocycle due to strong orbital hybridization. In such cases, the concept of a spatially localized chemical bond becomes ill-defined from a vibrational near-field perspective, and the model naturally converges to a distributed representation of bonding.

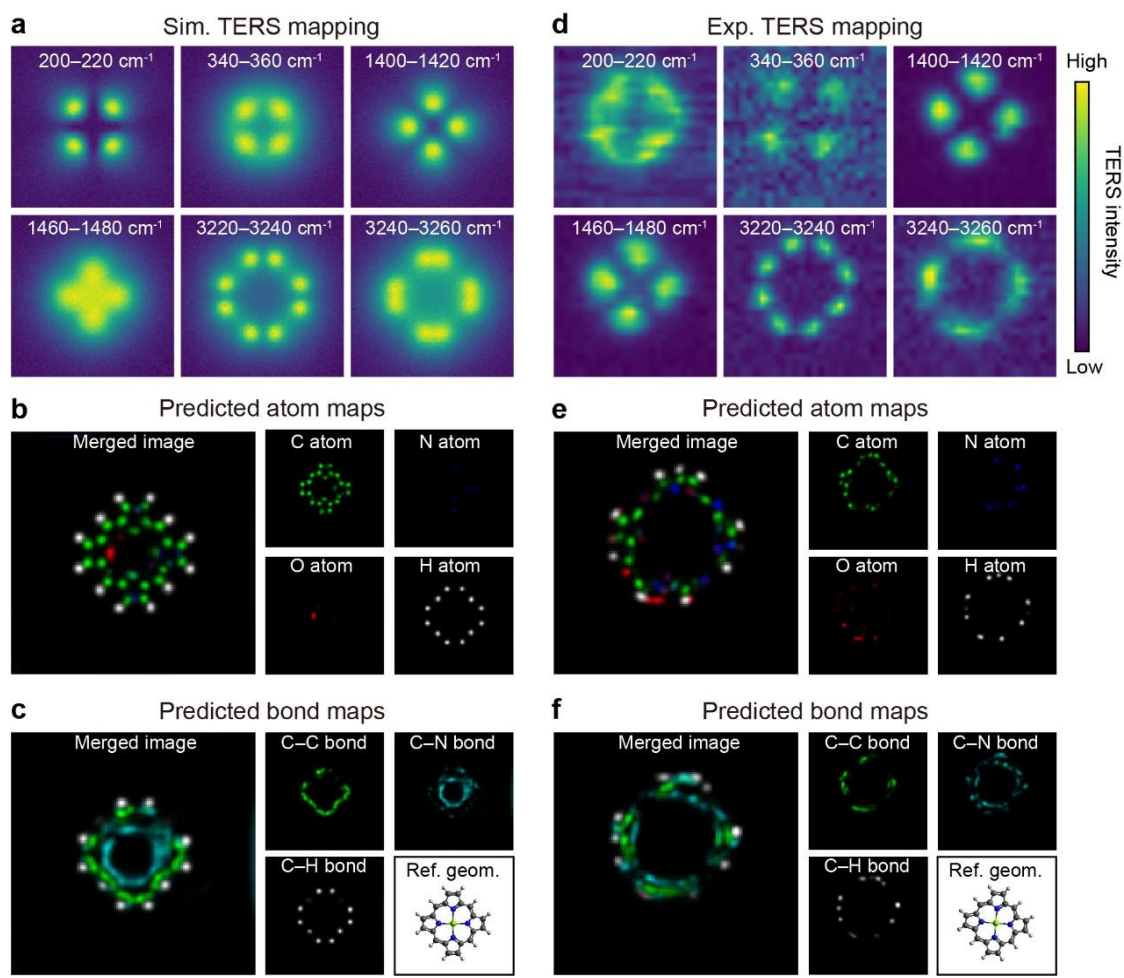

**Fig. 6. Identification of atoms and chemical bonds in a MgP molecule from the simulated and experimental TERS mappings. a−c**, Results obtained from simulated TERS data. **a**, Representative simulated TERS maps in six characteristic wavenumber regions. **b**, Atom probability maps predicted by ANet, shown as a merged image together with individual elemental channels. **c**, Bond probability maps corresponding to the top-three predicted bond types from BNet, shown as a merged image together with individual bond channels. The reference molecular geometry of MgP is shown in the lower-right corner, with orientation consistent with panels in (**a−c**). **d−f,** Results obtained from experimental TERS data. **d**, Six representative experimental TERS maps. **e**, Atom probability maps predicted by ANet from the experimental input, shown as a merged image together with individual elemental channels. **f**, Bond probability maps corresponding to the top-three predicted bond types from BNet. The reference molecular geometry is shown in the lower-right corner with orientation aligned to panels in (**d−f**). Size of image: 2.5 × 2.5 nm$^2$.

We then applied TERS-ABNet to experimentally measured TERS mappings of the MgP molecule. The raw experimental data consist of 25 × 25 pixel images spanning an area of 2.5 × 2.5 nm$^2$. To meet the network input requirements (128 × 128 pixels over the same area), the data were first denoised and subsequently unsampled via linear interpolation (Fig. 6d). Despite the substantially lower native resolution and the presence of experimental noise, the model retained partial atomic-level predictive capability. In particular, hydrogen atoms and several peripheral carbon atoms were consistently identified. The resulting carbon-probability distribution reproduces key structural motifs of the macrocycle, including the overall porphyrin ring and the characteristic five-membered pyrrole subunits.

Nevertheless, the model does not recover all atomic species, preventing full reconstruction of the molecular geometry at true atomic resolution. A major factor underlying this limitation is the spatial mismatch between the experimental TERS mapping and the real geometry of the molecule. Compared with the simulated TERS maps (Fig. 6a), the experimental maps (Fig. 6d) exhibit a more spatially extended signal profile, which reduces sensitivity to atoms located near the molecular center. In addition, the delocalized electronic structure of the conjugated macrocycle further blurs the correspondence between local vibrational response and individual atomic sites.

Despite these challenges, BNet successfully identifies segments of chemically meaningful connectivity, including portions of conjugated C−H and C−C bonds (Fig. 6f). This result demonstrates that even under non-ideal experimental conditions, structured deep learning models can extract useful atomistic constraints from experimental imaging data. Taken together, these findings highlight both the promise and the current limitations of data-driven molecular geometry reconstruction from experimental TERS measurements. For large and highly conjugated molecular systems, the combined effects of limited spatial resolution, vibrational and electronic delocalization, and experimental misalignment impose intrinsic bounds on achievable structural fidelity. Overcoming these challenges will require advances in experimental TERS imaging resolution, more diverse training datasets incorporating realistic experimental artifacts, and reconstruction algorithms capable of explicitly accounting for bond delocalization and partial observability in spectro-microscopic measurements.

## Conclusion

In this work, we establish TERS-ABNet as a unified deep-learning framework to address the inverse problem of molecular structure determination from tip-enhanced Raman spectroscopy. By learning the correspondence between spatially resolved vibrational fingerprints and chemical structures, TERS-ABNet directly infers atomic positions, chemical identities and bonding connectivity from TERS mapping data, without reliance on expert-driven spectral assignment or predefined chemical rules. This capability allows reliable recovery of molecular frameworks with atomic precision and extends to chemically complex motifs and nonplanar molecular geometries through transfer learning. A key finding of this work is that meaningful atomistic structural information can be computationally retrieved even from TERS mappings acquired at moderate spatial resolution. This relaxes the conventional requirement for Ångström-level near-field confinement, thereby lowering experimental barriers and improving the practical accessibility of automated structure determination in nanoscale spectroscopy. Validation using experimental measurements of MgP molecule further demonstrates the feasibility of extracting chemically consistent partial structural information under realistic conditions, while also highlighting current limits imposed by resolution, signal delocalization and measurement imperfections. More broadly, TERS-ABNet establishes a data-driven paradigm for inferring molecular structure from high-dimensional spectroscopic imaging, pointing toward increasingly automated and quantitative single-molecule characterization as experimental techniques continue to advance.

## Methods

### TERS-ABNet architecture

TERS-ABNet comprises two dedicated sub-networks: an atom prediction network (ANet) and a bond prediction network (BNet). Both networks are implemented within a 2D attention U-Net framework[34,35] (see Supplementary Figs. 1 and 2), utilizing an encoder–decoder paradigm tailored for high-precision semantic segmentation of spectroscopic images.

The encoder consists of three sequential 2D convolutional blocks with 3×3 kernels, interspersed with 2×2 average-pooling layers that progressively down-sample the spatial size from 128×128 to 16×16 pixels. This hierarchical contraction allows the model to capture long-range spatial dependencies across the molecular framework. Simultaneously, the feature-channel depth is refined from 160 to 64, facilitating feature abstraction of vibrational modes and the decoupling of interdependent spectral signatures. At the 16×16 latent resolution, a bottleneck module consisting of dual convolutional layers extracts high-level semantic representations essential for structural identification.

The decoder symmetrically reverses this structure through three up-sampling stages. Each stage integrates 2D convolutional layers with attention gates that selectively recalibrate features from the encoder via skip connections, thereby preserving fine-grained spatial information. After progressive refinement and channel alignment, the final output layer restores predictions to the original 128×128 grids. ANet produces a 4-channel atomic probability map corresponding to different elemental classes, whereas BNet outputs a 9-channel bond-type probability map, providing an explicit probabilistic representation of molecular geometry derived directly from TERS mappings.

**Molecular database**

The molecular dataset is primarily curated from the QM9 database[41], which contains small organic molecules with up to nine heavy atoms. From this repository, we select structures composed exclusively of C, N, O, and H atoms and exhibiting quasi-planar geometries. This planarity is quantitatively defined by the maximum out-of-plane height variation ($\Delta z$, measured between extremal atomic coordinates perpendicular to the best-fit molecular plane): < 0.1 nm for heavy atoms and < 0.16 nm for hydrogen atoms. This stringent geometric criterion yields 4,573 eligible planar structures. To improve the model's scalability toward larger molecular systems, we construct an auxiliary dataset from the FG26 database, containing molecules with up to 26 heavy atoms. Applying the same elemental and planarity constraints but restricting selection to molecules with more than ten heavy atoms resulted in 1,248 additional molecular structures. The combined dataset therefore comprised 5,821 planar molecules, which forms the foundation for the initial training and evaluation of TERS-ABNet.

To further extend the model to nonplanar molecular systems via transfer learning, we curate an additional dataset of 9,668 molecules (8,345 from QM9 and 1,323 from FG26). In this dataset, hydrogen atoms were allowed to deviate by 0.16–0.18 nm from the molecular plane, while the 0.10 nm threshold for heavy atoms was retained. This criterion is designed to accommodate out-of-plane functional groups and mild stereochemical distortions without introducing fully three-dimensional molecular frameworks.

Both the planar and nonplanar datasets are partitioned into training and testing subsets using a 9:1 ratio based strictly on unique molecular identities. Rotation-based data augmentation is performed only after dataset splitting to ensure that different orientations of the same molecule never appeared in both subsets, thereby preventing data leakage. Following this strict separation, TERS simulations are conducted to generate the final spectroscopic input maps used for model training, as described below.

**Simulation of TERS mapping**

TERS spectra are simulated following the methodology established in our previous work[13,33], integrating first-principles vibrational analysis with electromagnetic near-field modeling. Molecular geometries are first optimized, and vibrational properties are calculated using Gaussian16[42]. For molecules from the QM9 dataset, calculations are performed at the B3LYP/6-31G(2df,p) level, whereas larger molecules from the FG26 dataset are calculated using B3LYP/aug-cc-pVDZ. For each molecule, vibrational frequencies, normal-mode eigenvectors, and polarizability derivatives are extracted for subsequent TERS intensity simulations. The electromagnetic near-field distribution is simulated using MNPBEM toolbox[43]. The metallic tip is modeled with a spherical apex of 1 nm radius and 5 nm height, incorporating a 0.2 nm atomistic protrusion at its apex. The tip–substrate gap is fixed at 0.5 nm, and the molecular plane is positioned 0.3 nm above the Ag substrate.

TERS mappings are generated by scanning the tip over a 128×128 grid covering a 2.5×2.5 nm² area to fully encompass each molecule. At every grid point, the simulated spectrum is discretized into 20 cm$^{-1}$ bins spanning the 0–2000 cm$^{-1}$ and 2800–4000 cm$^{-1}$ regions. This procedure produces a 160-channel hyperspectral image stack for each molecule, yielding a structured 128 × 128 × 160 input tensor for TERS-ABNet that encapsulates the spatially resolved vibrational response.

**Atom and bond maps**

The spatial coordinates and chemical identities of atoms and bonds are encoded as multi-channel 2D probability maps with a size of 128×128 pixels. Atomic information is represented by a 4-channel tensor, with each channel corresponding to one elemental identity: carbon (C), nitrogen (N), oxygen (O), or hydrogen (H). Chemical connectivity is encoded as a 9-channel bond tensor, in which each channel represents one predefined bond type. To transform discrete structural annotations into continuous spatial representations compatible with convolutional learning, each atom or bond is modeled as a Gaussian intensity distribution centered at its true coordinate. The probability density at spatial position $\mathbf{r}$ is defined as $I(\mathbf{r}) = e^{-\alpha|\mathbf{r}-\mathbf{r}_0|^2}$, where $\mathbf{r}_0$ denotes the center coordinate of the atom or bond, and $\alpha$ controls the spatial spread of the distribution. For both atoms and chemical bond channels, $\alpha$ is set to 0.05 nm$^{-2}$, yielding localized yet spatially continuous probability fields that preserve positional precision while enabling stable network training.

**Data augmentation**

To enforce rotational equivariance and eliminate orientation-induced bias in TERS-ABNet, we adopted a systematic rotation-based data augmentation strategy. Each TERS hyperspectral mapping and its corresponding atom and bond probability maps are rotated by 90°, 180°, and 270°, increasing the planar training dataset from 5,821 to 23,234 samples.

For the transfer learning stage, class imbalance between the QM9 (8,345 molecules) and FG26 (1,323 molecules) subsets is addressed through targeted augmentation of the smaller FG26 cohort only. Applying identical rotational transformations results in a balanced transfer learning dataset comprising 13,637 samples.

To emulate experimental measurement variability and improve robustness against noise-induced degradation, additive Gaussian noise with a standard deviation of 0.02 (relative intensity units) is introduced to the simulated TERS mappings during training, approximating realistic signal-to-noise fluctuations observed in experimental TERS data and enhancing generalization to real-world data.

**Training hyperparameters**

TERS-ABNet is optimized using the Adam optimizer[44] as implemented in PyTorch[45]. The training process is performed with a mini-batch size of 32 samples per iteration. The initial learning rate is set to 0.0003 and decays multiplicatively by a factor of 0.92 every 10 epochs over a total training duration of 100 epochs. Model training employs the mean squared error (MSE, also referred as L2 loss) as the objective function, minimizing the discrepancy between the predicted probability maps and the corresponding ground-truth molecular structures.

**Transfer learning**

To extend TERS-ABNet to nonplanar molecular systems, we construct a dedicated transfer learning dataset comprising molecules containing methyl and ethyl functional groups, which introduce out-of-plane structural motifs and additional spectral complexity. The output layer of BNet was expanded from 9 to 11 channels to accommodate the newly introduced methyl and ethyl group probability maps. In contrast, the architecture of ANet remained unchanged, as the elemental categories (C, N, O, H) are consistent with those in the planar training set. The transfer-learning model is initialized using weights pre-trained on the planar molecular dataset. This initialization preserves low-level spectroscopic and spatial features learned from planar systems while enabling efficient adaptation to nonplanar geometries.

## Data availability

The data that support the findings of this study are available in the paper and its Supplementary Information.

## Code availability

The code repository for this work is available.

## Acknowledgements

This work was supported by the National Natural Science Foundation of China (Grant Nos. 22572182, 21333010, 11327805, 21790352, and 12334018), the National Key R&D Program of China (Grant Nos. 2021YFA1500500, 2024YFA1208100 and 2016YFA0200601), the Strategic Priority Research Program of Chinese Academy of Sciences (Grant Nos. XDB36000000), CAS Project for Young Scientists in Basic Research (YSBR-054), Anhui Initiative in Quantum Information Technologies (Grant

No. AHY090000), and Innovation Program for Quantum Science and Technology (Grant No. 2021ZD0303301). Theoretical simulations and model training were performed on the robotic AI-Scientist platform of Chinese Academy of Sciences.

## Author contributions

Yao.Z. and Z.C.D. conceived and supervised the project. J.C. and Yao.Z. designed the TERS-ABNet architecture and performed theoretical derivation and simulation. J.C. implemented the model framework, developed the code and performed model training. J.C., Yao.Z., Y.L. and Z.C.D. contributed to the data interpretation and co-wrote the manuscript. All authors discussed the results and commented on the manuscript.

## Competing interests

The authors declare no competing interests

## Additional information

**Supplementary information** The online version contains supplementary material available at https://doi.org/xxxxxxx.